# Optimization of p$^+$-ZnTe layer for Back Contacts of ZnTe Thin Film Solar Cells

Kyoung Su Lee, Gyujin Oh, and Eun Kyu Kim$^*$

*Department of Physics and Research Institute for Natural Sciences, Hanyang University,*

*Seoul 04763, Republic of Korea*

We have studied performance of ZnTe-based heterojunction diodes grown on p-GaAs substrates by pulsed laser deposition (PLD). In X-ray diffraction measurement, ZnTe and CdS thin films grown on p-GaAs (100) substrates appeared cubic crystalline structure, while ZnO film showed a hexagonal structure. The structure of n$^+$-ZnO/i-ZnTe/p-GaAs with a insertion layer of CdS buffer layer between n$^+$-ZnO and i-ZnTe showed a strong rectifying diode characteristics, and then its open circuit voltage ($V_{oc}$) and short circuit current density ($J_{sc}$) under 1,000 Wm$^{-2}$ air mass 1.5 global (AM 1.5G) were measured about 0.32 V and 0.61 mAcm$^{-2}$, respectively. By using another insertion of ZnTe:Cu layer as a p$^+$-layer for the contacts on p-GaAs substrate, $V_{oc}$ and $J_{sc}$ of n$^+$-ZnO/CdS/i-ZnTe/ZnTe:Cu/p-GaAs increased to 0.39 V and 0.81 mA/cm$^2$. As thickness of i-ZnTe increased from 210 to 420 nm, the photoelectric conversion efficiency of n$^+$-ZnO/CdS/i-ZnTe/ZnTe:Cu/p-GaAs increased to 0.24 %.

# 1. INTRODUCTION

II–VI compounds are widely used in solid-state devices such as mid-infrared lasers, photovoltaic cells, optical tweezers, and quantum cascade detector [1-4]. They are also important semiconductors for opto-electronic devices, photoconductive and photo-electric devices. ZnTe has, in particular, been studied for solar cell applications because of its optimum energy gap of 2.26 eV at room temperature and low electron affinity of 3.53 eV. ZnTe can be used for solar cells, for example, as a back surface field (BSF) layer and p-type semiconductor material for a CdTe/ZnTe structure [5-7]. ZnTe thin films can be grown by several techniques including pulsed laser deposition (PLD), molecular beam epitaxy, sputtering, and thermal evaporation [8-11]. Because of ZnTe generally shows p-type behavior, it is difficult to achieve n-type ZnTe for fabricating ZnTe homojunction. However, ZnTe heterojunctions can be made by using n-type ZnO layers [12]. There are challenges to fabricate ZnO/p-ZnTe heterojunction due to the different crystal structure and large lattice mismatch [13]. CdS can offer a compatible cubic n-type material with p-ZnTe and has a small lattice mismatch of 3.5% with ZnTe. Therefore, it can be possible to fabricate ZnO/CdS/ZnTe heterojunction diodes which are analogous to polycrystalline ZnO/CdS/CdTe heteronjunction diodes [14].

In this work, the ZnTe, ZnO, and CdS thin films were grown on p-GaAs substrates by PLD system, and their structural properties were measured by X-ray diffraction (XRD) and atomic force microscope (AFM). Performance of ZnTe-based heterojunctions incorporating CdS buffer layer and ZnTe:Cu BSF layers were analyzed by current density-voltage (J-V) characteristics.

# 2. EXPERIMETAL DESCRIPTION

All films for fabricating ZnTe-based solar cells were deposited on (100) p-GaAs substrates with a concentration of $5 \times 10^{17}$ cm$^{-3}$ by PLD. Before the film deposition, the p-GaAs substrate was cleaned by using chemical solutions of trichloroethylene, acetone, methanol and distilled water for 5 min each in an ultrasonic cleaner. A pulsed (10 Hz) Nd:YAG laser operating in a wavelength of 266 nm was used for ablating ZnTe, CdS, and ZnO targets. The plasma plume was created by illuminating the focused laser pulse onto the target at an angle of 55° with normal direction of the target. We focused the laser spot on the surface of the target by using an optical lens and then the focused spot size was approximately 1 mm$^2$. The calculated laser power density illuminating the target is about 3 J/cm$^2$. The targets were 6 mm thick, 30 mm in diameter, and composed of ZnTe, ZnTe doped with 2 wt.% Cu (ZnTe:Cu), CdS, and ZnO (99.999% purity). The distance from the target to the substrate was 50 mm and the speed of both rotating target and substrate was 5 rpm. The base pressure of the chamber was kept at a pressure around 8 x 10$^{-7}$ Torr by using a turbo-molecular pump.

The structural properties of thin films grown on p-GaAs substrates were analyzed by θ-2θ scan of XRD measurements (Rigaku SmartLab) using a Cu-Ka (1.5406A°) x-ray source. To measure thickness of ZnTe, CdS, and ZnO thin films, a line was drawn on c-plane sapphire substrate by using oily pens. And then, the thin films were grown on the substrate by PLD. After that, the line was removed by dipping them in acetone for 5 min. Through this process, surface height equal to thin film thickness was formed, and then the surface height was measured by using AFM (Park system, XEP-100). The scan area was 1 μm×1 μm with a resolution of 512×512 pixels. Newport solar simulator was used to provide a light illumination of AM 1.5G that was calibrated using a certified reference Si solar cell. Under dark and AM 1.5G, J-V curves were recorded by means of a Keithley model 2636-A digital source meter. All of the devices area were 1 cm$^2$.

## 3. RESULTS AND DISCUSSION

Figure 1 shows schematic diagram of n$^+$-ZnO/CdS/i-ZnTe/ZnTe:Cu/p-GaAs solar cell by PLD. Structural details of ZnTe-based solar cells were summarized in table 1. Here, the thickness of all films were measured by AFM. The ohmic contact of bottom p-GaAs side was made by evaporation of Ni/Au with thickness of 20 nm and 80 nm, respectively, and the grid pattern was formed using Al metal with thickness of 100 nm on the surface of n$^+$-ZnO layer.

Figure 2(a) and (b) show the surface morphologies of top n$^+$-ZnO layer over a scale 1 μm x 1 μm from the observed AFM images of device A and B, respectively. Upon inspection of the images in AFM measuring system, it was observed that the grain sizes for n$^+$-ZnO thin film of device A and B were about 120 and 180 nm, respectively. The root-mean-square (RMS) roughness of device A and B were about 2.3 and 7.8 nm, respectively. The PL intensity and PL decay rate of bound-exciton states of ZnO thin films increases with increasing grain size [14]. Because grain size of device B with 100 nm-thick CdS buffer layer was larger than that of device A without CdS buffer layer, PL intensity and PL decay rate of bound-exciton states of ZnO thin films improved by the CdS buffer layer between ZnO and ZnTe thin film.

Figure 3 shows XRD pattern of ZnTe, CdS, and ZnO thin films which were grown on p-GaAs (100) substrate. The XRD patterns of ZnTe thin film exhibited diffraction peaks at 2θ values of 25.30º, 29.22º, 49.44º, and 60.61º are correspond to (111), (200), (311), and (400) planes, respectively, of a polycrystalline zinc-blende structure. The XRD peak of CdS thin film at 2θ of 26.32° shows the preferential orientation along a (111) cubic plane. Because primary diffraction lines of hexagonal and cubic CdS thin film are in close proximity to each other, it was difficult to determine whether structure of CdS is hexagonal or cubic. Generally, the presence of (100) and (101) planes in XRD pattern of CdS indicates that structure of CdS is hexagonal, otherwise cubic [15]. Because the XRD pattern of CdS thin film did not contain (100) and (101) planes, the structure of CdS thin film was considered to be cubic. For ZnO thin film, two main diffraction patterns appeared at 2θ of 4.02° and 71.69°, which are corresponding to (002) and (004) planes of ZnO thin film, respectively. The peaks of (002) and (004) indicated that ZnO thin film had a preferred orientation with the c-axis perpendicular to the substrate and the structure was hexagonal wurtzite [16].

Figure 4 (a) and (b) show J-V characteristics of the device A, B, C, and D under dark and AM 1.5G illumination. The J-V characteristics were analyzed by using a modified diode-equation [17].

$$J = J_0 \exp\left[\frac{q}{\eta kT}(V - RJ)\right] + GV - J_L. \qquad (1)$$

In this equation, η is the diode ideality factor, q is the electron charge, k is Boltzmann's constant, T is temperature, $J_L$ is the light current, R is series resistance, G is conductance, and $J_o$ is the reverse saturation current. By applying equation (1) to the J-V data presented in these figures, the values for R, G, η, and $J_o$ can be extracted for all 4 devices both in the dark and under illumination, respectively. The results of this analysis for all 4 devices are summarized in table 2.

Under dark, device A showed large reverse saturation current and did not have characteristics of p-n diode possessing rectifying J-V behavior. However, device B demonstrated strong rectifying J–V behavior. $R_{dark}/G_{dark}$ of device A and B were 12Ω·cm²/13mS·cm⁻² and 25Ω·cm²/0.1 mS·cm⁻², respectively. $G_{dark}$ of device B was 130 times lower than that of device A while $R_{dark}$ of device B was about two times higher than that of device A. Under AM 1.5G illumination, $V_{oc}/J_{sc}$ of device A was 0.01 V/0.41 mA and $V_{oc}/J_{sc}$ of device B was 0.32 V/0.61 mA, so that PCE of device B (0.07 %) was seven times higher than that of device D (0.01 %). Here, the lattice mismatches of n⁺-ZnO/ZnTe and CdS/ZnTe were 14.2 and 3.5%, respectively. Because of these large lattice mismatches as well as different crystal structures between n⁺-ZnO (hexagonal)/ZnTe (cubic), the recombination of minority carrier at the n⁺-ZnO/ZnTe interface can be much stronger. However, the lattice mismatch of CdS/ZnTe was lower than that of n⁺-ZnO/ZnTe, and their crystal structure of CdS (cubic) and ZnTe(cubic) correspond to each other, so that recombining minority carrier at the

n$^+$-ZnO/ZnTe can be reduced. Therefore, the improvement of $V_{oc}$/$J_{sc}$ of device B can be ascribed to 100-nm-thick-CdS buffer layer which was inserted between n$^+$-ZnO and i-ZnTe.

Futher improvement of $V_{oc}$ and $J_{sc}$ appeared when a 140-nm-thick-ZnTe:Cu BSF layer was inserted between i-ZnTe and p-GaAs substrate. As comparing to device B without ZnTe:Cu BSF layer, $V_{oc}$/$J_{sc}$ of device C with ZnTe:Cu BSF layer increased to 0.39 V/0.81 mA cm$^{-2}$ and then PCE increased to 0.13 %. ZnTe:Cu layer of device C has a role of a BSF to repel minority carriers (electrons) at the interface between i-ZnTe and p-GaAs substrate [18]. Also, the carrier recombination at the interface between i-ZnTe and p-GaAs substrate can be reduced because of the interfacial barrier formation between i-ZnTe and the ZnTe:Cu BSF layers [19]. Therefore, the electrons can be reflected at this interface and they can be collected with a higher probability.

From the J-V characteristics under dark condition, the values of η and $G_{dark}$ of device D with 420-nm-thick i-ZnTe film were 3.3 and 0.05 mS/cm2, respectively, which were the lowest values, while $R_{dark}$ was the highest in the devices. Under light illumination for the device structures with n$^+$-ZnO/CdS/i-ZnTe/ZnTe:Cu/p-GaAs, it appeared that the photovoltaic property was dependent on the thickness of i-ZnTe absorber layer. We observed an increase in $J_{sc}$ (1.32 mA/cm$^2$) with increasing thickness of i-ZnTe film. The enhanced $J_{sc}$ can be attributed to the increased light absorption and charge carrier extraction of i-ZnTe film. As total film thickness of the devices increased from 360 (device A) to 790 nm (device D), $G_{dark}$/$G_{light}$ decreased from 13 mS·cm$^{-2}$/27 mS·cm$^{-2}$ to 0.05 mS·cm$^{-2}$/1.1 mS·cm$^{-2}$ and $R_{s,dark}$/$R_{s,light}$ increased from 12Ω·cm$^2$/8Ω·cm$^2$ to 12Ω·cm$^2$/8Ω·cm$^2$. Consequently, solar cell parameters of ZnTe based solar cells were improved by insertion of CdS buffer layer on i-ZnTe and ZnTe:Cu BSF layer on p-GaAs.

## 4. CONCLUSION

In this work, ZnTe-based heterojunction diodes were grown on p-GaAs substrates by PLD.

From XRD measurements, ZnTe and CdS thin films showed cubic structure but ZnO thin film appeared hexagonal structure. N$^+$-ZnO/i-ZnTe/p-GaAs device showed large reverse saturation current and low $V_{oc}$ (0.01 V) and $J_{sc}$ (0.41 mA/cm$^2$) due to different crystal structure and large lattice mismatch of 12.5% between ZnO and ZnTe. However, the n$^+$-ZnO/CdS/i-ZnTe/p-GaAs device demonstrated strong rectifying J-V behavior and $V_{oc}$ and $J_{sc}$ and PCE of the device increased to 0.32 V, 0.6 mA/cm$^2$, and 0.07 % which means that the insertion of CdS buffer layer on ZnTe has a major role in preventing recombination of minority carrier at n$^+$-ZnO/i-ZnTe interface. $V_{oc}$, $J_{sc}$, and PCE of n$^+$-ZnO/CdS/i-ZnTe(210nm)/ZnTe:Cu/p-GaAs device was 0.39 V, 0.81 mA/cm$^2$, and 0.13 %, respectively. The increased $V_{oc}$, $J_{sc}$, and PCE were attributed to insertion of ZnTe:Cu layer on p-GaAs substrate. As comparing of n$^+$-ZnO/CdS/i-ZnTe (210nm)/ZnTe:Cu/p-GaAs device, $J_{sc}$ and PCE of n$^+$-ZnO/CdS/i-ZnTe(420nm)/ZnTe:Cu device increased to 1.32 mA/cm$^2$ and 0.24 %. It was found that the improved $J_{sc}$ and PCE of the device can be attributed to the increased thickness of i-ZnTe (420 nm).

## ACKNOWLEDGEMENTS

This work was supported by the National Research Foundation of Korea (NRF) grant funded by the Korea government (MSIP) (NRF-2014R1A2A1A11053936).

# Table and figure captions

Table 1. Film thickness of ZnTe-base solar cell structures.

Table 2. Solar cell parameters of the four devices under dark and AM 1.5G illumination.

Figure 1. Schematic diagram of $n^+$-ZnO/CdS/i-ZnTe/ZnTe:Cu/p-GaAs solar cell.

Figure 2. Surface morphology of (a) $n^+$-ZnO thin film of device A and (b) device B measured by AFM

Figure 3 XRD patterns of ZnTe, CdS, and ZnTe grown on p-GaAs substrate by PLD

Figure 4. J-V curves of device A, B, C, and D under the dark (a) and (b) AM 1.5G illumination

Table 1. Film thickness of ZnTe-base solar cell structures

| Device | Approximate layer thickness (nm) | | | | |
| --- | --- | --- | --- | --- | --- |
| | $n^+$-ZnO | n-CdS | i-ZnTe | ZnTe:Cu | Total thickness |
| A | 150 | - | 210 | - | 360 |
| B | 150 | 100 | 210 | - | 460 |
| C | 150 | 100 | 210 | 120 | 580 |
| D | 150 | 100 | 420 | 120 | 790 |

Table 2. Solar cell parameters of the four devices under dark and AM 1.5G illumination

| Device | $J_{sc}$ (mA/cm²) | $V_{oc}$ (V) | FF (%) | PCE (%) | $R_{s,dark}$ (Ω cm²) | $R_{s,light}$ (Ω cm²) | $G_{dark}$ (mS cm⁻²) | $G_{light}$ (mS cm⁻²) | $n_{dark}$ | $n_{light}$ |
|---|---|---|---|---|---|---|---|---|---|---|
| A | 0.41 | 0.01 | 20 | 0.01 | 12 | 8 | 13 | 27 | - | - |
| B | 0.61 | 0.32 | 36 | 0.07 | 25 | 21 | 0.1 | 0.7 | 4.9 | 3.9 |
| C | 0.81 | 0.39 | 40 | 0.13 | 59 | 25 | 0.1 | 0.8 | 3.8 | 2.6 |
| D | 1.32 | 0.39 | 43 | 0.24 | 74 | 35 | 0.05 | 1.1 | 3.3 | 2.3 |

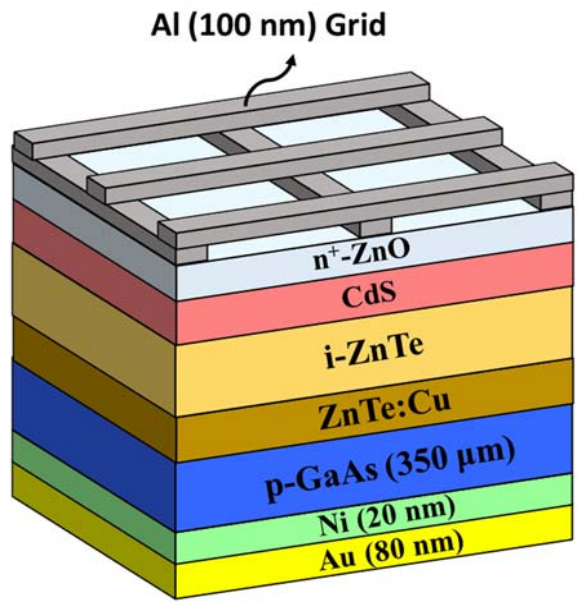

Figure 1 Kyoung Su Lee, *et al*.

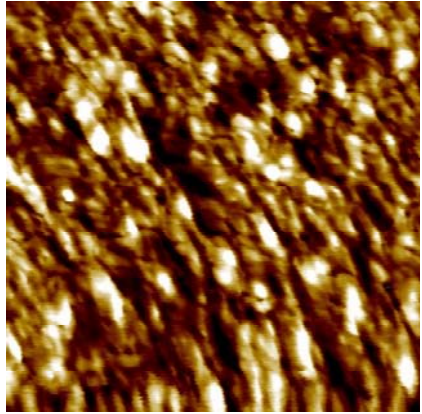

(a)

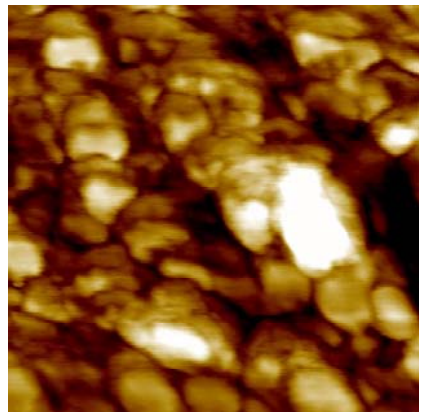

(b)

Figure 2 Kyoung Su Lee, *et al*.

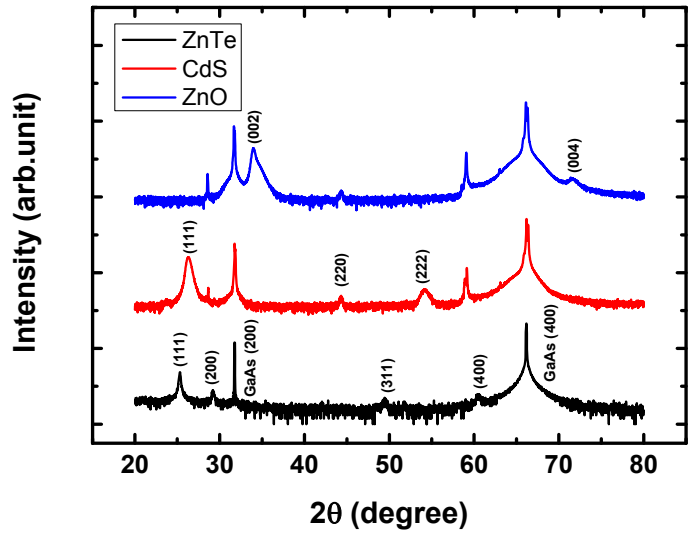

Figure 3 Kyoung Su Lee, *et al*.

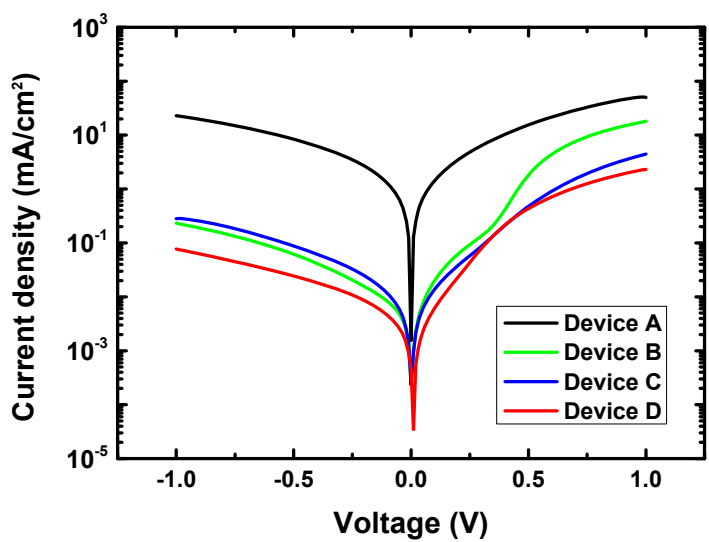

(a)

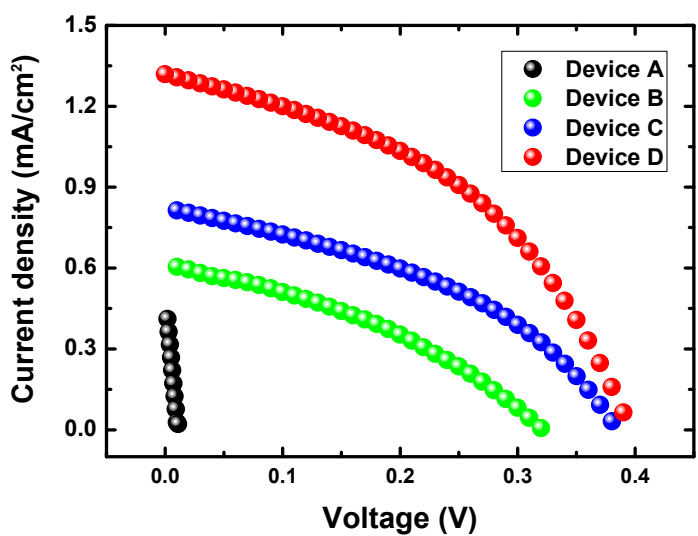

(b)

Figure 4 Kyoung Su Lee, *et al*.